\documentclass[%
 reprint,
 amsmath,amssymb,
 aps,
]{revtex4-2}

\preprint{APS/123-QED}

\usepackage{graphicx}
\usepackage{dcolumn}
\usepackage{bm}
\usepackage{xcolor}
\usepackage{sistyle}
\SIthousandsep{,}

\begin{document}

\preprint{APS/123-QED}

\title{Spectroscopy of the Hyperfine Structure of HD$^+$ in Rotationally Excited States\\with 10-ppb Uncertainty}

\author{D. Kliukin}
\altaffiliation[Present address ]{ASML, Veldhoven, The Netherlands}
\author{V. Barb\'e}
\altaffiliation[Present address ]{Laboratoire Kastler Brossel, Sorbonne Universit\'e, CNRS, ENS-PSL Research University, Coll\`ege de France, 4 place Jussieu, 75005 Paris, France}
\affiliation{LaserLaB, Department of Physics and Astronomy, Vrije Universiteit Amsterdam, De Boelelaan
	1100, 1081 HZ Amsterdam, The Netherlands}
\author{{J.-Ph. Karr}}
\affiliation{Laboratoire Kastler Brossel, Sorbonne Universit\'e, CNRS, ENS-PSL Research University, Coll\`ege de France, 4 place Jussieu, 75005 Paris, France}
\author{T. Klijn Velderman}
\author{J.B.E. Schokking}
\affiliation{LaserLaB, Department of Physics and Astronomy, Vrije Universiteit Amsterdam, De Boelelaan
	1100, 1081 HZ Amsterdam, The Netherlands}
\author{J. John}
\altaffiliation[Present address ]{Science Institute and Faculty of Physical Sciences, University of Iceland, 107 Reykjavík, Iceland}

\author{K.S.E. Eikema}
\author{J.C.J.Koelemeij}
\email[Corresponding author: ]{j.c.j.koelemeij@vu.nl}
\affiliation{LaserLaB, Department of Physics and Astronomy, Vrije Universiteit Amsterdam, De Boelelaan
	1100, 1081 HZ Amsterdam, The Netherlands}
\hyphenation{ro-vibrational}

\date{\today}

\begin{abstract}
We report the first measurement of the total hyperfine interval in the manifold of the ($v=0,L=3$) rovibrational state of HD$^+$ using microwave spectroscopy of HD$^+$ ions in a linear Paul trap. To overcome the low (0.2\%) occupancy of the quantum states involved, we employ a novel technique that combines intermittent Majorana depolarization of the HD$^+$ ensemble with redistribution of rotational-state population by blackbody radiation to effectively amplify the occupancy by a factor of 25. This enables the observation of a single field-insensitive magnetic subcomponent of the hyperfine transition, leading to a measured total hyperfine interval of \num{1\,050\,804.511}$\pm 0.011$\,kHz with an unprecedented relative uncertainty of 10 parts per billion. This differs from the theoretically predicted value \num{1\,050\,802.78}$\pm 0.89$\,kHz by 1.9$\sigma$. We show that our experimental value is consistent with the recently measured hyperfine structure of the ($v=0,L=0$) rovibrational state of HD$^+$ [C.~M. K{\"o}nig \textit{et al.}, High-precision Penning trap spectroscopy of the ground state spin structure of HD$^+$, Phys. Rev. Lett. \textbf{136},143002 (2026)], for which a similar deviation from theory was found. Our work may help resolve puzzling discrepancies between the theoretical and experimental hyperfine structure observed previously in optical rovibrational spectra of HD$^+$.
\end{abstract}

\maketitle

The spectrum of molecular hydrogen ions (MHIs) can be accurately determined using both theoretical and experimental methods, making them a sensitive probe of fundamental physics with features that are complementary to those of atoms and elementary particles~\cite{Schiller02102022}. The comparison between theoretical and measured MHI transition frequencies allows testing the validity of molecular \textit{ab-initio} quantum electrodynamics (mQED) calculations, or extraction of the values of the fundamental constants involved in these calculations~\cite{Karr2016,Karr2025}. MHI transitions are particularly sensitive to fundamental mass ratios, such as the proton-to-electron mass ratio ($m_\text{p}$/$m_\text{e}$), and can also probe the Rydberg constant ($R_{\infty}$) and the charge radii of the MHI constituents, such as the proton charge radius ($r_\text{p}$)~\cite{Karr2016,Karr2025,Schiller2024}. 
%

Rovibrational laser spectroscopy of the hydrogen deuteride ion (HD$^+$) has proven particularly successful in recent years. The availability of dipole-allowed, narrow transitions in HD$^+$ has opened the way to a series of precision measurements of its rotational and vibrational transition frequencies~\cite{Patra2020,Alighanbari2020,Kortunov2021,Alighanbari2023}. In combination with results from Penning-trap mass spectrometry and spin-precession spectroscopy~\cite{Fink2021,Rau2020,Heisse2019,Sturm2014}, this has enabled improved determinations of $m_\text{p}$/$m_\text{e}$ and of the relative atomic electron mass, $A_r$(e)~\cite{Karr2023}. As a result, HD$^+$ spectroscopy now contributes to the CODATA recommended values of the fundamental constants (CODATA 2022)~\cite{Mohr2025}. In addition, HD$^+$ spectroscopy has enabled a variety of searches for physics beyond the standard model, and provided some of the most stringent tests of mQED~\cite{Salumbides2013,Biesheuvel2016,Alighanbari2020,Patra2020,Germann2021,Kortunov2021,Alighanbari2023,Delaunay2023}. 

\begin{figure}
	\includegraphics[width=0.95\linewidth]{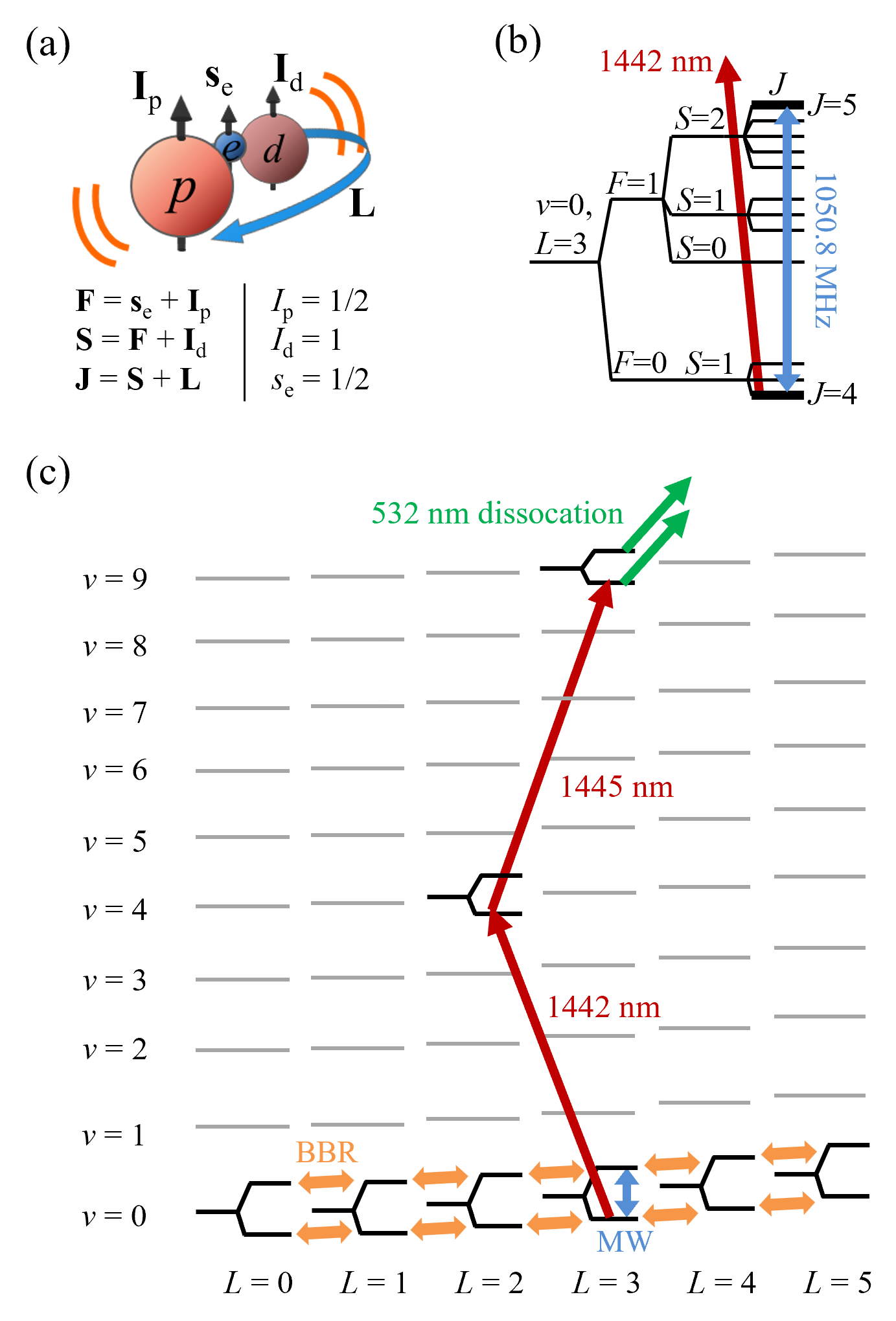}
	\caption{\label{levelDiagram} (a) Schematic depiction of the HD$^+$ molecular ion and angular momentum coupling scheme. (b) Partial energy level diagram of the hyperfine structure in the ($v=0,L=3$) rovibrational state (for a mathematical description, see~\cite{Karr2023}). The bold outer levels indicate the total hyperfine interval that is addressed by MW radiation and REMPD lasers that are resonant only with the $J=4$ hyperfine level. (c) Partial hyperfine-rovibrational level diagram (for each $L$, only the outer hyperfine levels are shown) and full spectroscopic scheme. Blackbody radiation and spontaneous emission drive rotational transitions on a $1\,$s-$100\,$s time scale~\cite{Tran2013}, which replenishes the population of the $J=4$ hyperfine state in $v=0,L=3$ while it is being dissociated through REMPD. This leads to enhanced REMPD loss of HD$^+$. If the microwave drive is resonant with the $J=5\rightarrow J=4$ transition, additional HD$^+$ ions are transferred from $J=5$ to $J=4$, and higher losses are observed.}
\end{figure}

Despite these successes, discrepancies between the measured and calculated HD$^+$ transition frequencies have emerged which are not yet understood. In particular, the $(F,S,J)=(0,1,4)$ and $(F,S,J)=(1,2,5)$ hyperfine components of the $(v=0,L=3)\rightarrow (v=9,L=3)$ vibrational transition, measured in our group using precision laser spectroscopy~\cite{Patra2020}, deviate from their theoretical values by 6.9\,$\sigma$ and 6.7\,$\sigma$, respectively~\cite{Karr2023}. Similar discrepancies have been observed for two hyperfine components of a rotational transition in HD$^+$, measured by other researchers~\cite{Alighanbari2020,Alighanbari2023,Karr2023}. In the above notation and throughout the rest of this work we follow the labeling of quantum numbers of Karr and Koelemeij~\cite{Karr2023,SM}, and the angular momentum coupling scheme shown in Fig.~\ref{levelDiagram}a.

These discrepancies may be the result of overlooked terms in the current hyperfine structure calculations or overlooked systematic effects in the measurements. However, the fact that reported direct hyperfine structure measurements in non-deuterated H$_2^+$ are consistent with theoretical predictions~\cite{Jefferts1969,Haidar2022} points to deuteron-specific physics and/or symmetry breaking as possible origin of the discrepancies. Here, it is worth noting that a $2.6\sigma$ tension between theory and experiment was observed in the hyperfine structure of muonic deuterium~\cite{Bonilla2026}, while good agreement was found for muonic hydrogen~\cite{Kalinowski2018}. Although the origin of this tension may be different from the cause of the discrepancies in HD$^+$, the present hyperfine issues in small deuterated atomic and molecular systems warrant further investigation. 

The HD$^+$ hyperfine discrepancies also affect the HD$^+$ input data for the 2022~CODATA adjustment, where expansion factors were introduced to artificially increase the experimental and theoretical error bars in order to reduce the impact of the discrepancies~\cite{Karr2023,Mohr2025}. This expansion factor currently does not limit the constants obtained from the 2022~CODATA adjustment, as the HD$^+$ theoretical uncertainty is dominated by that of the spin-averaged rovobrational transition frequencies. However, it may become a limiting factor if the rovibrational theory is improved, and may reduce the utility of HD$^+$ data for improved determinations of $m_\text{p}$/$m_\text{e}$,  $R_{\infty}$, and r$_\text{p}$~\cite{Karr2025,Schiller2024}. 

This situation calls for additional, more direct measurements of the hyperfine structure of HD$^+$. A recent investigation of the $v=0,L=0$ hyperfine structure of a single HD$^+$ ion, stored in a cryogenic Penning trap, yielded accurate experimental values of the spin coefficients $E_4(v,L)$ and $E_5(v,L)$, which characterize the strength of the proton-electron and deuteron-electron Fermi-contact interactions, respectively~\cite{Koenig2026}. With fractional uncertainties of 44 and 151 parts-per-billion (ppb),  these represent an improvement over the uncertainty of theoretical predictions of $E_4(0,0)$ and $E_5(0,0)$ by factors of 20 and 4, respectively~\cite{Haidar2022}. However, the experimental value $E^\text{exp}_4(0,0)$ was also found to be 1.9$\sigma$ larger, and $E^\text{exp}_5(0,0)$ 3.1$\sigma$ larger than the theoretical prediction, indicating a tension between theory and experiment. 

In this Letter, we present direct microwave (MW) spectroscopy of the $(F,S,J)=(1,2,5)\rightarrow (F,S,J)=(0,1,4)$ transition (hereafter denoted as the $J=5\rightarrow J=4$ transition) in the manifold of the $v=0,L=3$ rovibrational state of HD$^+$, which is relevant to the CODATA adjustment~\cite{Karr2023,Mohr2025}. Achieving a fractional uncertainty of 10~ppb, the smallest for any HD$^+$ hyperfine feature observed so far and a factor of 81 smaller than the present theoretical uncertainty, our measurement forms a benchmark for mQED theory, and a critical piece of information toward solving the hyperfine discrepancies. 
\begin{figure}[!t]
\includegraphics[width=0.4\textwidth]{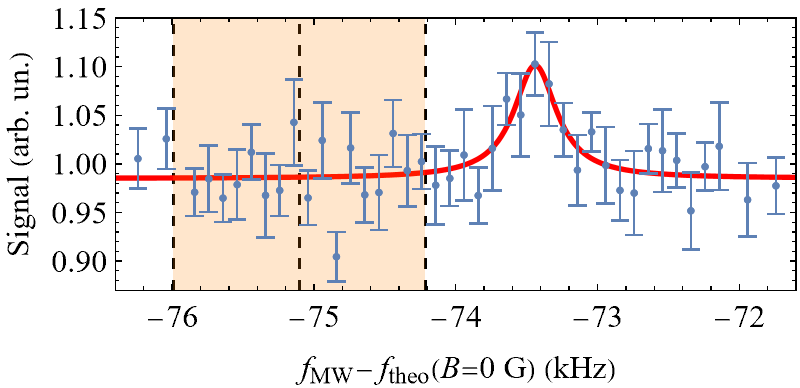}
	\caption{\label{scans} Spectrum of the $J=5,M_J=0\rightarrow J=4,M_J=-1$ transition acquired at 100-Hz step size and bias magnetic field of 1.99\,G (blue data points). Each data point represents the loss of HD$^+$ (in arbitrary units) averaged over 90 30-s experimental sequences, with  error bars indicating the standard error of the mean. The red curve is an unweighted nonlinear least-squares fit of a Lorentzian line shape to the data. The mQED theoretical prediction and its standard uncertainty are indicated by the light orange shaded area and dashed lines.}
\end{figure}

The experimental setup is similar to that described in our previous works~\cite{Patra2020,thesisSayanPatra}, with the addition of a flexible broadband antenna emitting a MW magnetic field at 1\,050.8\,MHz to drive hyperfine transitions in HD$^+$ (Fig.~\ref{levelDiagram}b). We load a few thousand beryllium ions (Be$^+$) in a linear Paul trap, which are laser cooled by continuous-wave (CW) 313~nm laser radiation obtained from a fiber-laser based system~\cite{Wilson2011}, leading to the formation of a millikelvin-cold Coulomb crystal. HD$^+$ ions are produced by electron-impact ionization and accumulate inside the crystal, where they are sympathetically cooled by the surrounding Be$^+$ ions. Typically $\sim$\,50\,-\,100 HD$^+$ ions are obtained during the loading procedure, all of which decay within 1~s to the vibrational ground state ($v=0$) by spontaneous emission. During spectroscopy, we apply a bias magnetic field of 1--2\,G along the trap axis using a pair of Helmholtz coils, in combination with two orthogonal pairs of shim coils. 

Partial energy level diagrams pertaining to relevant states of the HD$^+$ ion are shown in Fig.~\ref{levelDiagram}. The HD$^+$ ions are initially distributed among the six lowest rotational levels ($L=0\,,...\,,\,5$), following a room-temperature thermal distribution~\cite{Koelemeij2007,Patra2020}. About 20\,$\%$ of the ions are in the spectroscopically relevant $L=3$ state, which includes a total of 12 hyperfine states and 84 magnetic substates. For detecting MW transitions in HD$^+$, two narrow-linewidth CW external cavity diode lasers (ECDLs) at 1442\,nm and 1445\,nm excite the $(v=0,L=3)\rightarrow (v=9,L=3)$ two-photon vibrational transition in HD$^+$, in combination with a 532\,nm CW laser that dissociates the vibrationally excited HD$^+$ ions. This dissociation process is known as resonance-enhanced multiphoton dissociation (REMPD) and is state-selective, as we tune the wavelength of the ECDLs to address the spectrally resolved $(F,S,J)=(0,1,4$) hyperfine line of the two-photon transition (Figs.~\ref{levelDiagram}b,c). The resulting loss of HD$^+$ ions is estimated by means of the secular excitation method~\cite{Roth2006,Biesheuvel2017}, in order to detect the $J=5\rightarrow J=4$ MW transition. 

The spectroscopic measurement sequence starts with loading a fresh sample of HD$^+$ ions, followed by an estimation of the number of HD$^+$ ions~\cite{SM}. Next, the population in the $J=4$ target state is depleted by REMPD as described above (Fig~\ref{levelDiagram}b,c). The ions initially present in $J=4$ are therefore lost. As explained in more detail below, we then attempt to drive the $J=5\rightarrow J=4$ hyperfine transition using a series of MW pulses near the resonance frequency of $1\,050.8$\,MHz, after which we deplete the $J=4$ manifold again via REMPD, and estimate the remaining number of HD$^+$ ions. If the microwave radiation is resonant with the $J=5\rightarrow J=4$ transition, HD$^+$ ions are transferred from $J=5$ to $J=4$, from where they are dissociated. This leads to an increase in HD$^+$ loss as compared to the case of non-resonant MW radiation. 

The HD$^+$  $J=5\rightarrow J=4$ transition at $1\,050.8$\,MHz is probed using MW pulses of 50\,ms duration, during which the near-reasonant 1442-nm laser is switched off to avoid AC-Stark shifts, followed by 50\,ms of REMPD with the MW field switched off~\cite{SM}. The 313-nm, 532-nm and 1445-nm lasers, which are not resonant with any transition connecting to the hyperfine levels in $v=0,L=3$, are constantly on during the experiment. This two-pulse procedure is repeated nine times (lasting 900\,ms in total), and is subsequently followed by 100\,ms of Majorana depolarization (see below). This 1-s sequence is repeated 30 times to interrogate one HD$^+$ sample repeatedly at the same MW frequency. Afterwards, the number of remaining HD$^+$ ions is estimated, the remaining ions are removed from the trap, and the MW frequency is tuned to the next frequency value before the next sequence begins. Each sequence yields one data point for a spectrum. Spectra are between 1 and 4\,kHz wide, with groups of 30 to 90 data points per frequency, spaced by 50 or 100~Hz intervals. A 4-kHz wide spectrum that encompasses both the theoretically expected line position and the experimentally observed line is shown in Fig.~\ref{scans}. Further details of the spectroscopic sequence can be found in the Supplemental Material~\cite{SM}. 
\begin{figure*}[!t]
\includegraphics[width=0.9\textwidth]{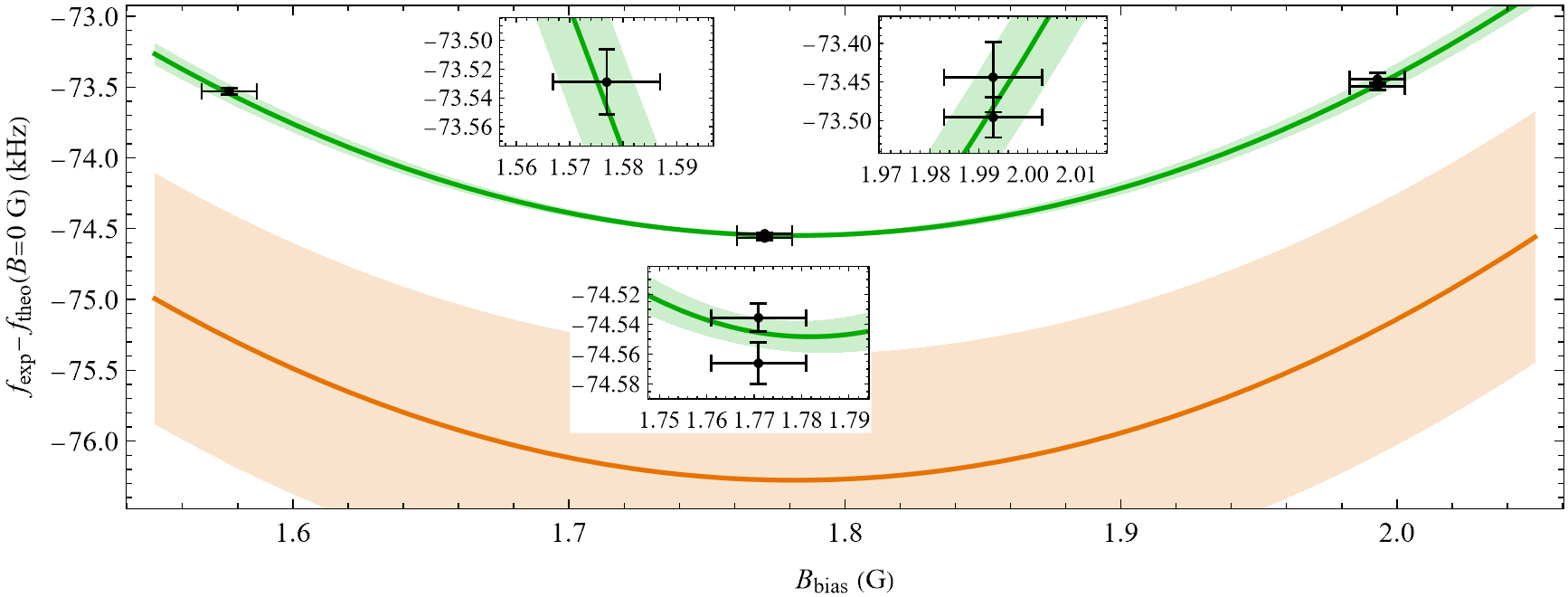}
	\caption{\label{fvsBplot} Overview of experimentally determined line centers as a function of bias field (black data points). The zero-field theoretical value $f_\text{theo}$ equals $1\,050\,802.79(89)$\,kHz. The insets zoom in on the various data points, with error bars representing standard uncertainty in the $(B_\text{bias},f_\text{exp})$ data points. The green curve is a weighted nonlinear least-squares fit (of the offset frequency and the bias magnetic field) of the theoretically computed transition frequency to the data, with the light-green shaded area indicating the 68\% confidence bands of the fit. The orange curve and light shaded area indicate the \textit{ab initio} theoretical prediction and its standard uncertainty, respectively.}
\end{figure*}

Control over Zeeman shifts~\cite{Bakalov2011} is critical in view of the unpaired electron in HD$^+$ which, given typical magnetic field variations and noise in our unshielded experimental apparatus of amplitude $\sim10\,$mG, may produce frequency shifts and broadening of up to 14\,kHz. This can be overcome by measuring the line $J=5,M_J=0\rightarrow J=4,M_J=-1$ (with $M_J$ the magnetic quantum number), which exhibits only a small quadratic Zeeman shift of 24.09\,kHz/G$^2$ for bias magnetic field values near 1.78\,G~\cite{SM}. However, due to the hyperfine multiplicities and the 300-K thermal rotational distribution, only 0.24\% of the HD$^+$ ions are initially in the desired $(v,L,F,S,J,M_J)=(0,3,1,2,5,0)$ initial state. This is about one order of magnitude smaller than is detectable in our experiment~\cite{Tran2013}. To improve the signal-to-noise ratio (SNR), we combine magnetic Majorana depolarization and blackbody-driven rotational thermalization, as explained in~\cite{SM}. We empirically find that this enhances the fraction of available HD$^+$ ions from 0.24\% to 6\% (i.e. a factor of 25 improvement), allowing the observation of the field-insensitive $J=5,M_J=0\rightarrow J=4,M_J=-1$ transition, on top of a typical baseline of 40\,-\,55\,$\%$ of REMPD-induced losses associated with other hyperfine states in the $F=0, S=1, J=2-4$ manifold. This novel method may be compared to existing strategies to increase population in single quantum states, such as optical pumping to the rotational ground state~\cite{schneider2010,Alighanbari2020}. An advantage of our approach is its wider applicability: it can in principle be used for any transition---microwave or optical---involving an initial rotational state with appreciable population at 300~K. In fact, it may facilitate high-resolution spectroscopy of transitions between single quantum states in other sympathetically cooled hydride-molecular ions for which REMPD is feasible.

To determine the in-situ bias magnetic field with high precision, we implemented coherent microwave spectroscopy at 1250\,MHz of the $(f=2,m_f=-2)\rightarrow (f=1,m_f=-1)$ hyperfine transition in the $^1$S$_{1/2}$ ground state of Be$^+$~\cite{Kunihiro1998}. The Zeeman effect of this transition is accurately known~\cite{Bollinger1983}, suggesting that this transition could be used to calibrate the magnetic field. However, this would  require a full characterization of systematic offsets in our Be$^+$ MW spectroscopy, and additional measures to ensure that the magnetic-field calibration remains valid also for the experimental conditions under which HD$^+$ spectroscopy is carried out. To circumvent this, we only exploit the $\pm 1$\,mG reproducibility of our Be$^+$ MW spectrosocopy to set the bias magnetic field with high precision, and calibrate it a posteriori using the theoretical Zeeman effect of HD$^+$ as explained below. The Zeeman effect of HD$^+$ is known from theory with an uncertainty of several parts in $10^8$~\cite{SM}, and can therefore be considered exact for the Zeeman shifts $\mathcal{O}(1\,\text{MHz})$ in our experiment. Over the course of days, we observe bias-field variations of 10--20\,mG, while measurements with fluxgate magnetic field sensors reveal fast noise with RMS amplitude 10\,mG. We assign a 10-mG statistical uncertainty to our magnetic field determination, which may be considered conservative given that slow bias-field variations are to a large extent detected and compensated with the help of the Be$^+$-derived magnetic field estimates, and the magnetic field noise will be (at least partially) averaged out over the 30-s acquisition time of one data point.

After acquisition of a spectrum, we check for the presence of a spectroscopic signal in 200-Hz frequency windows using a Welch's t-test that we developed previously~\cite{Patra2020,thesisSayanPatra}, and which helps differentiate signal from background under low-SNR conditions. We also correct for slow variations in the spectroscopic signal over time to improve the SNR and the sensitivity of the Welch's t-test~\cite{SM}. A MW spectrum recorded at a magnetic field of 1.99\,G is presented in Fig.~\ref{scans}. The spectrum is composed of three runs done for identical experimental settings, where the data for each run were acquired automatically over a span of 9 consecutive hours. This includes 5.3~hours of bare data acquisition (required for the 630 spectroscopy sequences of 30~s each) and overhead due to periodic reloading of HD$^+$, the time needed for checks that we implement for long-term stability of the setup such as monitoring and reloading of the Be$^+$ crystal when the ion number goes down, and other actions such as re-locking the frequency of the ECDLs after they unlock.

Figure~\ref{fvsBplot} shows the found line positions as a function of the magnetic field, and the fitted Zeeman shift and frequency offset to the $J=5,M_J=0\rightarrow J=4,M_J=-1$ transitions. Magnetic field values of the data points are initially estimated from Be$^+$ MW spectroscopy, but the fit to the theoretical curve allows adjustment of an overall offset magnetic field. The fit reveals an offset of $-10(6)$~mG relative to the (uncalibrated) Be$^+$-derived estimates of the bias field. The magnetic-field residuals of our data are consistent with the 10-mG statistical uncertainty (Fig.~\ref{fvsBplot}). From the fit we deduce a zero-field experimental transition frequency with an uncertainty of 8.8~Hz, which is subsequently corrected for various systematic effects to arrive at a transition frequency $f_\text{exp}=1\,050\,804.511(11)$\,kHz~\cite{SM}. Leading systematic effects (by uncertainty) are the AC-Zeeman shift associated with the RF currents in the ion-trap electrodes (5.4~Hz), the AC-Stark shift due to the 532~nm laser (3.0~Hz), the AC-Zeeman shift due to the MW probe field (1.8 Hz), and the quadratic Zeeman shift (1.2~Hz) caused by slow AC fields, i.e. fields with frequencies much smaller than relevant Zeeman and hyperfine shifts, such as those produced by the turbo molecular pumps. Other shifts such as the electric quadrupole shift, the AC-Stark shift due to the 1445~nm laser, and the propagation of uncertainty of the hyperfine theory that enters the Zeeman fit model contribute much less than 1~Hz of uncertainty~\cite{SM}. The found experimental value is 81 times more accurate than the theoretical prediction, $f_\text{theo}=1\,050\,802.78(89)$\,kHz, and $1.9\sigma$ higher in value~\cite{Korobov2020,Haidar2022}. 

It is worth mentioning that the $E_4(0,3)$ and $E_5(0,3)$ spin coefficients contribute 95\% of the total hyperfine interval. This suggests that our measurement, if combined with the values $E_4^\text{exp}(0,0)$ and $E_5^\text{exp}(0,0)$ reported by K{\"o}nig et al.~\cite{Koenig2026}, may be used to validate the rotational scaling of the theoretical spin coefficients $E_4$ and $E_5$. To this end, we multiply the ratio of theoretical coefficients $E_4(0,3)/E_4(0,0)$ with $E_4^\text{exp}(0,0)$ to obtain a hybrid experimental-theoretical estimate, $E_4^\text{hyb}(0,3)$. In a similar fashion, we obtain $E_5^\text{hyb}(0,3)$. We determine the uncertainty of the hybrid coefficients by propagating the uncertainties of $E_4^\text{exp}(0,0)$ and $E_4(0,3)/E_4(0,0)$, taking into account correlations~\cite{Alighanbari2020,Koelemeij2022,Haidar2022,Karr2023}. Importantly, because of the one-to-one correlation between uncertainties of $E_i(0,0)$ and $E_i(0,3)$ ($i=4,5$), the ratios $E_i(0,3)/E_i(0,0)$ have zero uncertainty. With the obtained values and uncertainties of $E_4^\text{hyb}(0,3)$ and $E_5^\text{hyb}(0,3)$, combined with the other hyperfine coefficients $E_j(0,3)$ (with $j=1$--$3,6$--$10$) from theory~\cite{Korobov2020,Haidar2022,Karr2023}, we obtain a hybrid prediction of the transition frequency $f_\text{hyb}=1\,050\,804.525(43)_\text{exp}(219)_\text{theo}$~kHz. The experimental uncertainty contribution stems primarily from the 41-Hz uncertainty of $E_4^\text{exp}(0,0)$~\cite{Koenig2026}
, while the theoretical uncertainty contribution is predominantly due to the uncertainty of $E_1(0,3)$~\cite{Haidar2022}. The combined uncertainty of $f_\text{hyb}$ is 224~Hz, which is four times smaller than the uncertainty of $f_\text{theo}$. Although the hypothesis of perfectly correlated uncertainties in $E_i(0,0)$ and $E_i(0,3)$ is well motivated (see the discussion in Appendix C of~\cite{Haidar2022}), it is not expected to be fully exact. It is therefore useful to study the impact of a slight deviation of the correlation coefficient from 1 to assess the robustness of our conclusions. In the (hypothetical) case that the correlation between $E_i(0,0)$ and $E_i(0,3)$ ($i=4,5$) would not equal 1 but 0.98 instead, the uncertainty of the hybrid value would increase to 279~Hz, still three times smaller than the uncertainty of $f_\text{theo}$.

$f_\text{hyb}$ deviates from $f_\text{exp}$ by 14~Hz, or 0.06$\sigma$. We point out that our single measurement of $f_\text{exp}$ cannot be used to determine  individual values of both $E_4^\text{exp}(0,3)$ and $E_5^\text{exp}(0,3)$; rather, it constrains them through the equation $f_\text{exp}=f_\text{theo}(E_4, E_5)$ (assuming fixed values of the other spin coefficients $E_j$). Nevertheless, the agreement between $f_\text{exp}$ and $f_\text{hyb}$ is consistent with the hypothesis that the present hyperfine theory of HD$^+$, while not yet fully in agreement with measured hyperfine structure, predicts the correct scaling with rotational quantum number $L$. Conversely, if one would start out from the assumption that the theory scales correctly with $L$, the above analysis shows that our measurement of the $v=0, L=3$ hyperfine splitting is consistent with the measurements of the $v=0, L=0$ hyperfine structure of~\cite{Koenig2026}.

In summary, we have performed microwave hyperfine spectroscopy of the $(F,S,J)=(1,2,4)\rightarrow (F,S,J)=(1,2,5)$ transition, corresponding to the total hyperfine interval of the $(v=0,L=3)$ rovibrational state of HD$^+$. With a fractional uncertainty of 10~ppb, our measurement is four times more accurate than a recent determination of the hyperfine structure of the $(v=0,L=0)$ state in HD$^+$ by K{\"o}nig \textit{et al.}~\cite{Koenig2026}, and 81 times more accurate than the theoretical prediction~\cite{Korobov2020,Haidar2022}. We observe a deviation of $1.9\sigma$ from theory, comparable to the tension with theory observed for the $(v=0,L=0)$ state~\cite{Koenig2026}. We furthermore exploit known correlations between theoretical hyperfine coefficients to scale the results by K{\"o}nig \textit{et al.} for $(v=0,L=0)$ to the case of $(v=0,L=3)$ with essentially zero added uncertainty. This enables a hybrid experimental-theoretical prediction of the hyperfine splitting in $(v=0,L=3)$ that can be directly compared with our measurement. The excellent agreement indicates that the theoretical hyperfine structure exhibits a rotational scaling of the proton-electron and deuteron-electron Fermi-contact interactions that is consistent with high-accuracy hyperfine spectroscopy. These insights may contribute to resolving an ongoing hyperfine puzzle in HD$^+$~\cite{Haidar2022,Karr2023}. For example, combined with the hyperfine splitting measured through two-photon spectroscopy of the $(v=0,L=3)\rightarrow (v=9,L=3)$ vibrational transition~\cite{Patra2020}, our measurement implies a significant deviation relative to theory of $-6.8(1.2)$~kHz for the total hyperfine interval in $v=9, L=3$~\cite{Haidar2022}, which we plan to measure directly in the future. 

This work was supported by the U.S. Department of Commerce, National Institute of Standards and Technology under financial assistance award number 60NANB21D184, and by the Netherlands Organisation for Scientific Research under grant number OCENW.M.21.141. The work of J.-Ph.K. was part of 23FUN04 COMOMET that has received funding from the European Partnership on Metrology, co-financed by the European Union's Horizon Europe Research and Innovation Program and by the Participating States. Funder ID: 10.13039/100019599.

The measurement data, fit model, and analysis that support the findings of this work are available from  \{public DataverseNL handle to be inserted by authors\}.
%
%

\bibliography{apssamp}

\end{document}